\documentclass[fleqn,twoside]{article}
\usepackage[headings]{espcrc2}

\readRCS
$Id: espcrc2.tex,v 1.2 2004/02/24 11:22:11 spepping Exp $
\usepackage{graphicx}
\usepackage[figuresright]{rotating}

\newcommand{\pythia}{P\scalebox{0.8}{YTHIA\ }}
\newcommand{\pythias}{P\scalebox{0.8}{YTHIA7\ }}
\newcommand{\pythiae}{P\scalebox{0.8}{YTHIA8\ }}
\newcommand{\amegic}{\scalebox{0.8}{AMEGIC++\ }}
\newcommand{\apacic}{\scalebox{0.8}{APACIC++\ }}
\newcommand{\herwig}{\scalebox{0.8}{HERWIG\ }}
\newcommand{\herwigpp}{\scalebox{0.8}{HERWIG++\ }}

\newcommand{\sherpa}{S\scalebox{0.8}{HERPA\ }}
\newcommand{\alpgen}{A\scalebox{0.8}{LPGEN\ }}
\newcommand{\comphep}{C\scalebox{0.8}{OMP}H\scalebox{0.8}{EP\ }}
\newcommand{\madevent}{M\scalebox{0.8}{AD}E\scalebox{0.8}{EVENT\ }}

\newcommand{\AmS}{{\protect\the\textfont2
  A\kern-.1667em\lower.5ex\hbox{M}\kern-.125emS}}

\title{Event generator for the LHC}
\author{T.~Gleisberg\address[TUD]{Institut f\"ur theoretische Physik, TU Dresden, D-01062 Dresden, Germany}, 
        S.~H{\"o}che\addressmark[TUD], 
        F.~Krauss\addressmark[TUD], 
        A.~Sch{\"a}licke\address[zeuthen]{
           DESY Zeuthen, Platanenallee 6, 15738 Zeuthen}, 
        S.~Schumann\addressmark[TUD], 
        J.~Winter\addressmark[TUD]}
       
\runtitle{Event generator for the LHC}
\runauthor{T.~Gleisberg et al.}

\begin{document}
\begin{abstract}
\noindent
In this contribution the new event generation framework \sherpa will be 
presented. It aims at the full simulation of events at current and future
high-energy experiments, in particular the LHC. Some results related to 
the production of jets at the Tevatron will be discussed.
\vspace{1pc}
\end{abstract}

\maketitle

\section{INTRODUCTION}

\noindent
The observation and interpretation of multi-particle, multi-jet final 
states will be in the centre of the physics programme at the LHC. They
serve as signals or backgrounds for new physics; as an example consider
the production and decay of heavy SUSY particles. This shift of focus 
towards higher multiplicities translates directly into new challenges 
for the simulation of such events and necessitates the construction of new 
simulation tools; for an overview on currently available tools cf.\
\cite{Dobbs:2004qw}.

\noindent
The multi-purpose event generator \sherpa \cite{Gleisberg:2003xi} 
is one of these new tools, 
which are currently under construction. Others, namely
\pythias \cite{Lonnblad:1998cq,Bertini:2000uh}, \pythiae \cite{Pythia8}
and \herwigpp \cite{Gieseke:2003hm} are complete rewrites of the 
well-established codes \pythia \cite{Sjostrand:2000wi,Sjostrand:2003wg} 
and \herwig \cite{Corcella:2000bw,Corcella:2002jc}, extending or 
improving their physics content. To exemplify this, consider the new 
parton showering algorithms \cite{Gieseke:2003rz,Sjostrand:2004ef}, 
that are or will be included. All these new codes are written in the 
object-oriented programming language C++; in the beginning it was planned 
that at least the successors of \pythia and \herwig would be based on 
the same underlying machinery and a repository of common classes, 
CLHEP \cite{Lonnblad:1994kt}, which is also used in experiments. 
However, this turned out to be not feasible, and \pythiae and \sherpa 
are relying on their own framework.

\section{PRESENTING SHERPA}
\noindent
It is fair to state that, at the moment, \sherpa is the one of the new 
simulation tools, which is most advanced when it comes to the ability to 
actually generating events. Currently, the following physics modules
are implemented:
\begin{itemize}
\item PDFs:\\
      Various PDFs - CTEQ \cite{Pumplin:2002vw} and 
      MRST \cite{Martin:1999ww} in their original form as well as many other 
      PDFs through the LHAPDF library \cite{LHAPDF} - are interfaced.
\item Matrix elements:\\
      \amegic \cite{Krauss:2001iv} is a matrix element generator to describe 
      hard scattering processes and decays at the tree level. Apart from the 
      full SM \amegic contains the full MSSM \cite{Haber:1984rc} in the notation
      of \cite{Rosiek:1989rs,Rosiek:1995kg} and an ADD model of large extra 
      dimensions \cite{Arkani-Hamed:1998rs} with its implementation described 
      in \cite{Gleisberg:2003ue}. SUSY particle spectra are provided 
      through the SUSY Les Houches accord interface \cite{Skands:2003cj}.

\item Parton showers:\\
      For multiple QCD bremsstrahlung, i.e.\ the emission of secondary partons, 
      \sherpa relies on \apacic \cite{Krauss:2005re}, which uses, similar to
      \pythia \cite{Sjostrand:2000wi}, an ordering by virtuality supplemented 
      with an explicit angular veto to ensure a proper treatment of quantum coherence.
      \noindent
      The merging of the hard matrix elements for multijet production
      and the subsequent parton shower is achieved according to the merging 
      procedure proposed in \cite{Catani:2001cc,Krauss:2002up} and implemented
      in \cite{Schalicke:2005nv}.
\item Multiple parton interactions:\\
      A first simulation of the ``hard'' underlying event in the spirit of 
      \cite{Sjostrand:1987su} but supplemented with parton showering has been 
      implemented and tested. A more involved model is currently in preparation.
\item Hadronisation:\\
      The translation of the emerging partons into primordial hadrons is taken 
      care of by the Lund string model \cite{Andersson:1998tv}. This, as
      well as subsequent hadron decays are realized by an interface to the 
      corresponding \pythia routines. 
      \noindent
      However, a new version of cluster fragmentation \cite{Winter:2003tt} is 
      ready to be fully implemented in the near future.
\end{itemize}

\section{RESULTS}

\subsection{Matrix elements}

\noindent
\amegic automatically constructs Feynman diagrams and helicity 
amplitudes \cite{Kleiss:1985yh,Hagiwara:1985yu} for a given set of processes. For 
the helicity amplitudes, the formulation of \cite{Ballestrero:1992dv} is employed.
Having constructed them, \amegic simplifies and combines them by factoring out
common parts and then writes the results out in library files to be compiled and
linked with the core program. This leads to a drastic reduction of computing time. 
For the Monte-Carlo integration over phase space, the multi-channel approach of 
\cite{Berends:1994pv,Kleiss:1994qy} is being used. For each Feynman diagram, suitable 
phase space mappings are constructed and also written out as library files.
During integration, the weight optimisation procedure selects successful channels
having a large impact on the integration. In \amegic, after a number of integration 
steps, these winning channels are then further optimised by employing VEGAS
\cite{Lepage:1980dq} to select random numbers for them.

\noindent
\amegic has exhaustively been tested for a large number of production cross 
sections for six-body final states at an $e^+e^-$-collider \cite{Gleisberg:2003bi} and 
various processes at the LHC, see \cite{MC4LHC}.
As an example for the latter, consider the processes $pp\to e^-\nu_e+n\mbox{\rm jets}+X$
and $pp\to e^-\nu_eb\bar b+n\mbox{\rm jets}+X$ at the LHC. Cross sections for 
these processes, obtained through \alpgen \cite{Mangano:2002ea}, \comphep 
\cite{Pukhov:1999gg}, \madevent \cite{Stelzer:1994ta,Maltoni:2002qb}, and \amegic 
can be found in Table \ref{Xsecs}
\footnote{ 
  For details of the calculational setup and more results, cf.\ the MC4LHC homepage 
  \cite{MC4LHC}.
}.
\begin{sidewaystable}
  \label{Xsecs}
  \begin{center}
    \begin{tabular*}{15cm}{@{\extracolsep{\fill}}|c|c|c|c|c|c|c|c|c|c|}
      \hline
      \multicolumn{2}{|c|}{ X-sects (pb)} & \multicolumn{7}{c|}{ Number of jets}\\\hline
      \multicolumn{2}{|c|}{ $ e^- \bar \nu_e $ + $n$ QCD jets }
      & 0 & 1 & 2 & 3 & 4  & 5 & 6\\
      \hline
      \multicolumn{2}{|c|}{\alpgen}   
      & 3904(6)& 1013(2) & 364(2)& 136(1) & 53.6(6) & 21.6(2) & 8.7(1)\\
      \multicolumn{2}{|c|}{\comphep}  
      & 3947.4(3)& 1022.4(5)& 364.4(4)& & & &\\
      \multicolumn{2}{|c|}{\madevent} 
      & 3902(5)& 1012(2)& 361(1)& 135.5(3) & 53.6(2) & &\\
      \multicolumn{2}{|c|}{\amegic}   
      & 3908(3) & 1011(2) & 362(1) & 137.5(5) & 54(1)  & &\\
      \hline
    \end{tabular*}\\[5mm]
    \begin{tabular*}{10cm}{@{\extracolsep{\fill}}|c|c|c|c|c|c|c|c|}
      \hline
      \multicolumn{2}{|c|}{ X-sects (pb)} & \multicolumn{5}{c|}{ Number of jets}\\\hline
      \multicolumn{2}{|c|}{ $ e^- \bar \nu_e $ + $b\bar b$ }& 0 & 1 & 2 & 3 & 4  \\
      \hline
      \multicolumn{2}{|c|}{Alpgen}   & 9.34(4)& 9.85(6)& 6.82(6)& 4.18(7)& 2.39(5) \\
      \multicolumn{2}{|c|}{CompHEP}  & 9.415(5)& 9.91(2)& & &  \\
      \multicolumn{2}{|c|}{MadEvent} & 9.32(3)& 9.74(1)& 6.80(2)&  & \\
      \multicolumn{2}{|c|}{Sherpa}   & 9.37(1) & 9.86(2) & 6.87(5) &   &   \\
      \hline
    \end{tabular*}
    \caption{
      Compilation of results for cross sections for
      $pp\to e^-\nu_e+n\mbox{\rm jets}+X$ and $pp\to e^-\nu_eb\bar b+n\mbox{\rm jets}+X$ 
      at the LHC.
    }
  \end{center}
\end{sidewaystable}

\noindent
In these comparisons, \amegic proved to work for up to eight external particles, 
and, thus, \sherpa includes a state-of-the-art matrix element generator, one
of the key elements of modern event generators. 

\subsection{Merging of matrix elements and the parton shower}

\noindent
In order to fully exploit the power of such multi-particle matrix elements,
they have to be combined with the parton shower which models subsequent,
secondary emission of softer QCD quanta. There are different ways to do so,
among them MC@NLO \cite{Frixione:2002ik}.
An alternative approach \cite{Catani:2001cc,Krauss:2002up} is to combine 
tree-level matrix elements for different jet multiplicities. This is done by
defining two disjoint regions of jet production and evolution, separated by
a jet measure defined according to the $k_\perp$ algorithm 
\cite{Catani:1991hj,Catani:1992zp,Catani:1993hr}. Then the matrix elements
are reweighted with suitable Sudakov form factors such that the corresponding
matrix element becomes ``exclusive'', and the parton showers are vetoed such 
that no extra jet is produced in the showering.
This approach guarantees independence of the jet separation definition at
leading logarithmic order. The algorithm has been implemented in full
generality in \sherpa \cite{Schalicke:2005nv}
\footnote{There exist some variations of this approach
\cite{Mangano:2001xp,Lonnblad:2001iq,Lavesson:2005xu,Mrenna:2003if}
with different technical realisations of the same idea. 
}, forming one of its cornerstones.
The approach and its implementation in \sherpa has been tested by comparing 
both with data and other codes, in a number of processes, cf.\ for example
\cite{Krauss:2004bs,Krauss:2005nu,Gleisberg:2005qq}. The findings in 
these comparisons were:
\begin{itemize}
\item Self-consistency:\\
  Varying the $k_\perp$ cut in the internal jet definition for the merging
  and the maximal number of jets taken care off by the matrix elements, the
  approach has been found to be extremely stable and independent of the jet
  definition.
\item Scale-independence:\\
  The shapes of characteristic distributions such as the transverse momentum
  of jets etc.\ are surprisingly stable (deviations of the order of 20\% and 
  less) under global variations of  the renormalisation and factorisation scale 
  in the matrix elements, Sudakov weights and the parton shower. The total cross
  section, being calculated at leading order only, however, depends much stronger
  on these choices.
\item Comparison with NLO results:\\
  The shapes of distributions obtained by \sherpa are in excellent agreement
  with those obtained by full NLO calculations.
\end{itemize}

\begin{figure}[htb]
  \begin{center}
    \vspace*{-1cm}
    \includegraphics[width=6cm]{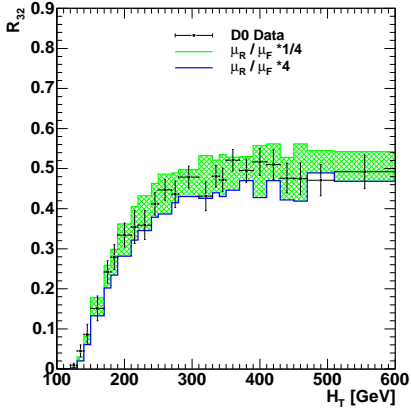} 
    \vspace*{-1cm}
    \caption{\label{R32_3}
      $R_{32}$ in dependence of $H_T$ for different global factors on
      the renormalisation and factorisation scale.
    }
  \end{center}
\end{figure}

\begin{figure}[htb]
  \begin{center}
    \vspace*{-1cm}
    \includegraphics[width=6cm]{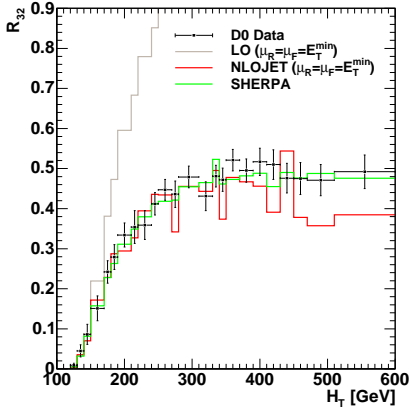} 
    \vspace*{-1cm}
    \caption{\label{R32_4}
      $R_{32}$ in dependence of $H_T$ as described by \textsc{Sherpa} and a
      full NLO calculation.}
  \end{center}
\end{figure}

\noindent
In this presentation, the merging approach will be applied to the case of
jet production at Tevatron, Runs I and II. At Run I, the D0 collaboration measured
the ratio of the three-to-two jet rate $R_{32}$, using the midpoint algorithm with
different $E_\perp$ of the jets and in different regions of pseudorapidity
of the jets \cite{Abbott:2000ua}. The stability of the results obtained by \sherpa
under variations of renormalisation and factorisation scales $\mu_{R,F}$ is 
demonstrated in Fig.\ \ref{R32_3}. In Fig.\ \ref{R32_4} the \sherpa results are 
contrasted with those of a full NLO calculation \cite{Nagy:2003tz}. In all these 
figures $R_{32}$ is plotted against $H_T$, the scalar sum  of the transverse momenta 
of all hard objects in the detector. Jets are defined by $E_\perp>40$ GeV, 
$|\eta|\le 3$ and $R=0.7$.
Again, stability of the results obtained by \sherpa under scale variations is found; the 
agreement with a full NLO calculation for the result is remarkable. These findings,
however, are a continuation of what has been found before for other processes.

\begin{figure}[htb]
  \begin{center}
    \vspace*{-1cm}
    \includegraphics[width=6cm]{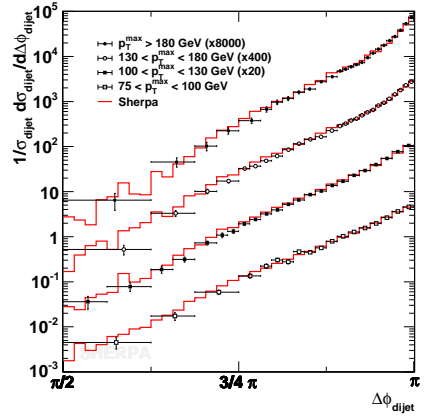} 
    \vspace*{-1cm}
    \caption{\label{Decorr}
      The azimuthal decorrelation of the two leading jets in the transversal plane.
      The four curves correspond to four different bins of $p_\perp$ of the
      leading jet.
    }
  \end{center}
\end{figure}

\noindent
At Run II, the D0 collaboration measured the angular decorrelation in the
azimuthal plane of the two leading jets in jet production \cite{Abazov:2004hm}. 
In Fig.\ \ref{Decorr} the results of \sherpa are contrasted with the experimental 
findings.
Again, \sherpa is capable of precisely reproducing the QCD radiation pattern 
in these events.

\section{CONCLUSIONS}

\noindent
In this publication the event generator \sherpa has been presented. Results have
been presented for the working of its internal matrix element generator \amegic,
a state-of-the art tool. The implementation of the merging procedure is a key ingredient of
the \sherpa event generator. Results for jet production at the Tevatron prove that
the merging of tree-level matrix elements and parton showers is work in a systematically
correct manner. Further tests, especially in the simulation of more processes, are ongoing.
The results obtained so far indicate that \sherpa is perfectly suitable to meet the enhanced
demands of the community to reliably simulate physics processes at the next generation of
collider experiments.

\section*{ACKNOWLEDGEMENTS}
\noindent
Financial support by BMBF, DESY and GSI is gratefully acknowledged.

\bibliographystyle{h-elsevier2}
\bibliography{sherpa}
\end{document}